\documentstyle[psfig]{mn}

\def\refitem{\par\parskip 0pt\noindent\hangindent 20pt}
\def\lesssim{\mathrel{\hbox{\rlap{\hbox{\lower4pt\hbox{$\sim$}}}\hbox{$<$}}}}
\def\gtrsim{\mathrel{\hbox{\rlap{\hbox{\lower4pt\hbox{$\sim$}}}\hbox{$>$}}}}

\def\BB{\hbox{\bf B}}
\def\VV{\hbox{\bf V}}

\title[A disk-wind model with correct crossing of all MHD 
critical surfaces]
{A disk-wind model with correct crossing of all MHD 
critical surfaces}

\author[N. Vlahakis et al.]
{N. Vlahakis$^{1, 2}$, K. Tsinganos$^1$, C. Sauty$^3$,
E. Trussoni$^4$\\\\ 
$^1$Department of Physics, University of Crete, P.O. Box 2208, GR-710 03 
Heraklion, Crete, Greece\\
$^2$Present address: The University of Chicago, The Enrico Fermi Institute,  
5640 S. Ellis Av., Chicago, IL 60637, USA\\ 
$^3$Observatoire de Paris - Universit\'e Paris 7, DAEC, F-92190 Meudon, France\\
$^4$Osservatorio Astronomico di Torino, Strada Osservatorio 20,
I-10025 Pino Torinese, Italy}
\date{Accepted 2000 May 23. Received 2000 April 13; in original form 1999 August 20}       
\begin{document} 
\maketitle 

\begin{abstract}

The classical Blandford \& Payne (1982) model for the magnetocentrifugal 
acceleration and collimation of a disk-wind is revisited and refined.   
In the original model, the gas is cold and the solution is everywhere 
subfast magnetosonic.
In the present model the plasma has a finite temperature and the 
self-consistent solution of the MHD equations starts with a subslow 
magnetosonic speed which subsequently crosses all critical points, at  
the slow magnetosonic, Alfv\'en and fast magnetosonic separatrix surfaces. 
The superfast magnetosonic solution thus satisfies MHD causality. 
Downstream of the fast magnetosonic critical point the poloidal 
streamlines overfocus towards the axis and the solution is terminated. 
The validity of the model to disk winds associated with young 
stellar objects is briefly discussed.  

\vskip 0.5 true cm

\end{abstract}


\section{Introduction}

Astrophysical jets are systematically associated with the presence of an 
underlying accretion disk, both observationally and theoretically 
(see K\"{o}nigl \& Pudritz 2000 for a recent review). In the 
case of protostellar objects, accretion disks are resolved by means of 
infrared and millimeter surveys and interferometric mappings down to 
scales of a few tens of AU. In the optical and the near infrared, HST  
high resolution images of disks in several jet sources have also been 
obtained (Padgett et al. 1999). With an apparent relation found between 
accretion and ejection in the form of a strong correlation between outflow 
signatures and  accretion diagnostics (see e.g. Cabrit et al. 1990, 
Cabrit \& Andr\'e  1991, Hartigan et al. 1995), stellar jets seem to be powered 
by the gravitational energy released in the accretion process. 

These facts and considerations have led several authors to develop models of 
disk winds. The pioneering work of Bardeen \& Berger (1978) on a hydrodynamic 
radially  
self-similar model of a hot galactic wind was generalized in the
seminal paper of Blandford \& Payne (1982, henceforth BP82) by including a 
rotating magnetic field. In particular, in BP82 it was shown that a cold 
plasma can be launched magneto-centrifugally from a Keplerian disk, 
similarly to a bead on a wire, provided that the magnetic field lines
are sufficiently inclined from the axis. 
Since then, steady and axisymmetric MHD models, 
self-similar in the radial direction, have been successfully analyzed and 
generalized in the literature (see e.g. Contopoulos \& Lovelace 1994, 
henceforth CL94, Li 1995, 1996, Ferreira 1997, Ostriker 1997, Vlahakis \& Tsinganos 
1998, henceforth VT98, Lery et al. 1999). 

A major problem is however still open on the validity of the various classes 
of radially self-similar solutions analyzed so far. Because, as it is well known  
since the original work of Weber \& Davis (1967) on the rotating magnetized 
solar wind in the equatorial region, acceptable outflowing solutions must 
cross smoothly all singularities related to the characteristic speeds of the MHD 
perturbations, i.e., the poloidal Alfv\'en velocity and the slow/fast 
magnetosonic
velocities. However, in radially self-similar equations the  critical 
points  are not found where the poloidal speed of the flow  is equal to  the 
characteristic velocities of these magnetosonic waves. In the cold model of BP82 the 
``modified'' fast magnetosonic critical point (where $t=1$ in the BP82 notation) 
is 
found downstream of the  position where the poloidal velocity of the wind is 
equal to the fast magnetosonic velocity. Subsequently it has been shown 
that this is a general property of the axisymmetric steady MHD equations: the 
singularities of the equations coincide with the positions of the limiting 
characteristics, or separatrices, within the hyperbolic domain of the 
governing equations (Bogovalov 1994, Tsinganos et al. 1996). 
In particular, Bogovalov (1994, 1996) pointed out the key role played  by the 
singularity occurring at the fast magnetosonic separatrix surface (FMSS). 
Namely, the asymptotic region of the jet is causally disconnected from the 
base of the flow, only for solutions that cross the critical point at the FMSS. 
This means that every terminal perturbation or shock does not affect the
outflow structure upstream of the  position of this critical point.
And, Tsinganos et al. (1996) have given several analytical examples where the 
true singularities of the equations do not coincide with the positions 
where the governing partial differential MHD equations change character from 
elliptic to hyperbolic and vice versa.  
For the sake of simplicity from now on we shall indicate by `fast/slow 
magnetosonic singularity', or in short 'modified fast/slow', the critical 
points at the FMSS/SMSS. 
 
It turns out that in none of the previous models of disk-winds a solution 
has been found to cross the FMSS. For example, Li (1995, 1996) and  
Ferreira (1997), starting  from the accretion disk, succeeded to cross  
the slow magnetosonic and the Alfv\'en 
ones, but downwind turning points were found where the solutions terminate.   
Such solutions can be connected to infinity only through a shock, as 
suggested by Gomez de Castro \& Pudritz (1993). 
However in this case,  as the wind velocity is subfast magnetosonic, a 
temporal evolution of the outflow is expected (Ouyed \& Pudritz 1997). 
 
Cylindrically collimated solutions were found by Ostriker (1997) for a 
cold plasma, integrating the MHD system upstream from infinity and crossing 
the Alfv\'en singularity, but always in the subfast magnetosonic regime. 
On the other hand, it has been shown that in collimated  winds  oscillations 
of streamlines are a common feature (Vlahakis \& Tsinganos 1997). 
It thus seems that cylindrically collimated solutions without oscillations 
correspond to a rather particular choice of parameters that completely  
suppresses such oscillations. A slight change in these parameters  
induces the onset of oscillations which increase in amplitude until the 
configuration is destroyed (Vlahakis 1998).  Since the Ostriker (1997) solutions 
are asymptotically subfast magnetosonic  they are likely to be sensitive to 
perturbations from the external medium, unlike solutions that really satisfy all 
the criticality conditions. Therefore, such solutions are likely to be 
structurally and topologically unstable (Vlahakis 1998).

However, it has been shown by Contopoulos (1995) that, in the restricted case 
of a purely toroidal magnetic field, a smooth crossing of the  FMSS 
is possible. On the other hand in such a case an  
asymptotically  cylindrically collimated  configuration is not found; in fact,  
a new transition to subfast magnetosonic velocities  
must occur anyway for radially self-similar winds.  The only way out is then 
to match the superfast magnetosonic solution with a shock which is in this case 
in the physically disconnected domain.

In the present study we extend the analysis of BP82, CL94 and Contopoulos (1995)
showing that an exact and simultaneous smooth crossing of \underbar{all}  
three MHD critical surfaces is possible.  In Sec. 2 we define the equations of 
the hot wind in the framework of a radially self-similar approach and outline 
the numerical technique.  In Sec. 3 we explore  the solution topologies in the 
region around and particularly downstream of the FMSS, where the solution 
terminates, while in Sec. 4 are shown the features of a few solutions crossing 
all three critical points with conditions similar to those of
BP82. Finally, in Sec. 5 we discuss the possible astrophysical applications
of these solutions to stellar jets, and summarize the main implications
of our results in comparison with previous ones obtained by other authors.

\section{Model description}

In order to establish notation, in this Section we give a brief derivation 
of radially self-similar disk-wind models 
with polytropic thermodynamics. The derivation is along the lines of a 
systematic 
method which unifies all self-similar MHD outflows and includes the BP82 model 
as the simplest case (VT98). 
   
\subsection{General definitions and self-similar assumption}

In steady $(\partial t = 0)$ and axisymmetric $(\partial \phi = 0)$ MHD, 
the poloidal components of the hydromagnetic field ($\BB, \VV$) 
are defined in terms of the magnetic flux function $A$ and mass to magnetic flux 
function $\Psi_{A}(A)$ in cylindrical ($z\,, \varpi \,,\phi$) or 
spherical ($r\,, \theta \,,\phi$) coordinates, as:
\begin{equation}
\label{BpVp}
\BB_p=\nabla  \times \frac {A \hat {\phi}} {\varpi } 
\,,\qquad
\VV_p =\frac{\Psi_{A}(A)} {4 \pi \rho} \BB_p
\,.
\end{equation}
The azimuthal components are defined in terms of the 
total specific angular momentum $L(A)$ and of the corotation frequency 
$\Omega (A)$, which are functions of $A$ (Tsinganos 1982):
\begin{equation}\label{LO}
L(A) = \varpi\left( V_{\phi} - \frac{B_{\phi}}{\Psi_A}\right) 
\,, \;\;\;
\Omega (A) = \frac{1}{\varpi}\left( V_{\phi} - 
M^2\frac{B_{\phi}}{\Psi_A}\right)\,, 
\end{equation} 
and of the poloidal Alfv\'en number $M$ :
\begin{equation}\label{LOO}
M = \sqrt{4 \pi \rho} \frac{V_p}{B_p} = \frac {\Psi_A}{\sqrt{4 \pi \rho}}
\,.
\end{equation}
TransAlfv\'enic flows require that, when $M=1$, $V_{\phi}$ and $B_{\phi}$ 
are finite, i.e.:
\begin{equation}\label{transalfvenic}
{L\over \Omega } = 
\varpi_{\alpha}^2 (A) \equiv \varpi_\star^2 \alpha 
\,,
\end{equation}
where $\varpi_\star$ is the Alfv\'en cylindrical radius (the Alfv\'en lever 
arm) along the reference field line $\alpha=1$, with  
the dimensionless variable $\alpha$ defined as some function of 
the magnetic flux function $A$  which can be reversed to give:
\begin{equation}\label{inver}
A= \frac{B_\star \varpi_\star^2}{2}{\cal A}\left(\alpha\right)
\,.
\end{equation}
where $B_\star$ is a constant with the dimensions of a magnetic field.

As shown in VT98, all existing classes of {\it radially 
self-similar} MHD solutions can be constructed by making the following two key 
assumptions: \\
(i) the Alfv\'en number $M$ is solely a function of $\theta$,
such that the Alfv\'en surface is conical:
\begin{equation}
\label{assumptionM}
M \equiv M(\theta)
\,, 
\end{equation}
\noindent 
(ii) the cylindrical distance $\varpi$ to the 
polar axis of some fieldline labeled by $\alpha$, normalized to its 
cylindrical distance $\varpi_{\alpha}$ at the Alfv\'en point is 
also solely a function of $\theta$:
\begin{equation}
\label{assumptionG}
G\left(\theta\right) \equiv \frac{\varpi}{\varpi_{\alpha}} 
\,.
\end{equation}
 
Following these two assumptions the set of MHD equations is reduced to a system
of three ordinary differential equations in $\theta$ for 
$M(\theta)$, $G(\theta)$ 
and the $\theta$ -dependence of the gas pressure
(see VT98 for details).

\subsection{Polytropic thermodynamics}

Depending on the assumptions on the free integrals $A(\alpha)$, 
$\Psi_A(\alpha)$, $L(\alpha)$ and $\Omega(\alpha)$, a few classes of radially
self-similar solutions exist (see VT98). For only 
two of these classes a polytropic 
relationship between the gas pressure and the density is admitted:
$P=Q(\alpha)\rho^{\gamma}$, where $Q(\alpha)$ is the specific entropy 
(the first two cases listed in Table 3 of VT98). In such 
a case $A \propto \alpha^{x/2}$, $\Psi_A \propto \alpha^{(x - 3/2)/2}$, 
$\Omega \propto \alpha^{-3/4}$,  $L \propto \alpha^{1/4}$, and the system of 
the MHD equations reduces to two first order differential equations for 
$M(\theta)$ and  $G(\theta)$, supplemented by the Bernoulli integral 
which also provides the variable $\psi(\theta)$, the angle between a particular 
poloidal fieldline and the cylindrical direction $\hat \varpi$ at the 
spherical angle $\theta$. 
Note that the parameter $x$ (with the same notation as in CL94, while in 
VT98 $x$ was denoted by $F$)  
governs the scaling of the magnetic field, while the rotation law is
assumed Keplerian. This particular class corresponds to the radially 
self-similar solutions analyzed in CL94 which contains as a special case 
the classical BP82 solution with $x=0.75$.  
  
The full expressions of ${\rm d} M^2 /{\rm d} \theta$,  ${\rm d} G^2 /{\rm d} 
\theta$ and $\psi(\theta$)  are given in the Appendix, Eqs. (A1) - (A3)
The expressions for the physical variables become then:
\begin{equation}\label{norm}
\frac{\rho}{\rho_\star}= \alpha^{x-3/2}\frac{1}{M^2}
\,,\qquad 
\frac{P}{P_\star}= \alpha^{x-2-\gamma \left(x-3/2\right)} 
\left(\displaystyle \frac{ \rho}{\rho_\star}\right)^{\gamma} 
\,,
\end{equation}
\begin{equation}\label{norm1}\nonumber
\frac{\BB_p}{B_\star} = 
- \alpha^{\frac{x}{2}-1} \frac{1}{G^2}
\displaystyle \frac{\sin \theta}{\cos \left(\psi + \theta \right) }
\left(\sin \psi \hat{z} + \cos\psi \hat{\varpi}\right)\,, 
\end{equation}
\begin{equation}\label{norm2}\nonumber
\frac{\VV_p}{V_\star}= 
- \alpha^{-1/4} \frac{M^2}{G^2}
\displaystyle \frac{\sin \theta}{\cos \left(\psi +\theta \right) }
\left(\sin \psi \hat{z} + \cos\psi \hat{\varpi}\right)\,,
\end{equation}
\begin{equation}\label{norm3}
\frac{B_{\phi}}{B_\star}=-\lambda  {\alpha}^{\frac{x}{2}-1}
\displaystyle \frac{1-G^2}{G\left(1-M^2 \right)}
\,, 
\end{equation}
\begin{equation}\label{norm4}
\frac{V_{\phi}}{V_\star}=\lambda  {\alpha}^{-1/4}
\displaystyle \frac{G^2-M^2}{G \left(1-M^2 \right)}
\,.
\end{equation}
 
\subsection{Parameters}

At the Alfv\'en radius $\varpi_\star$ along the reference field line $\alpha=1$, 
we denote by $P_\star$ and $\rho_\star$ the pressure and density, respectively.  
The magnitude of the poloidal magnetic field at this Alfv\'en point is 
$- B_\star \, {\sin \theta_\star}/{\cos \left(\psi_\star +\theta_\star \right) 
}$
while the corresponding poloidal Alfv\'en speed is 
$-V_\star \, {\sin \theta_\star}/{\cos \left( \psi_\star +\theta_\star \right)}$,
with $B_\star=\sqrt{4\pi\rho_\star}V_\star$. 

The expressions of the free integrals defined in Sec. 2.1 can now be written as:
\begin{equation}\label{APSI(a)}
A=\displaystyle \frac{B_{\star}\varpi_{\star}^2}{x} \alpha^{x/2}\,,
\qquad
\Psi_A^2=\displaystyle 4 \pi \rho_{\star} \alpha^{x-3/2}
\,,
\end{equation}
\begin{equation}\label{OL(a)}
\Omega=\displaystyle \lambda \frac{V_{\star}}{\varpi_{\star}}
\alpha^{-{3/4}}
\,,\qquad 
L=\lambda V_{\star} \varpi_{\star} \alpha^{{1/4}}
\,,
\end{equation}
\begin{equation}\label{OP(a)}
E=V^2_{\star} \epsilon \alpha^{-1/2}  
\,,\qquad
V^2_{\star}=\displaystyle \frac{{\cal G}{\cal M}}{{\varpi_{\star}} \kappa^2}
\,,\qquad 
P_{\star}= \displaystyle \mu \frac{B^2_{\star}}{8 \pi}
\,,
\end{equation}
where $E$ is the sum of the kinetic, enthalpy, gravitational and Poynting 
energy flux densities per unit of mass flux density,
\begin{equation}\label{E(a)}
E(\alpha) = \frac{V^2}{2} + \frac{\gamma}{\gamma - 1} \frac{P}{\rho} - 
\frac{{\cal G}{\cal M}}{r} - \frac{\Omega}{\Psi_A} r \sin \theta B_{\phi}  
\,,
\end{equation}
while ${\cal G}$ and ${\cal M}$ are the gravitational constant and the mass of 
the central body, respectively.   

The solution of the system of Eqs. (A1) - (A3) depends on the six parameters 
$x$, $\gamma$, $\kappa$, $\lambda$, $\epsilon$ and $\mu$, introduced in Eqs. 
(\ref{APSI(a)}) - (\ref{OP(a)}) (but see the discussion in Sect. 2.4.3 for the 
free parameters of the model).  
Note that we have used for the parameters a similar but not an identical 
notation with 
BP82, since it occurred  to us that it is better to choose a different 
normalization. However, in the following we shall outline for convenience 
the correspondence between our parameters and those in BP82.  
 
Let us first discuss the physical meaning of the above parameters. First, 
the exponent $x$ is equal to $3/4$ in BP82, while in Ferreira (1997) 
it is related to the ejection index $\xi = 2(x-3/4)$. 
This index $\xi$ is related to the accretion rate and to the mass flux in 
the wind if also the structure of the disk is assumed radially self-similar 
(see e.g. Ferreira 1997). Second, we remind that $\gamma$ is the usual 
polytropic index. Next,  the constant $\kappa$ is the Keplerian speed
at radius $\varpi_{\star}$ on the disk,
in units of $V_{\star}$, i.e., it is proportional to the ratio 
of the Keplerian speed to the poloidal flow speed at the Alfv\'en radial 
distance, $V_{\rm p, A}$, and is related to the corresponding constant 
$\kappa_{\rm BP}$ in BP82. Since $\kappa$ is also proportional to the mass to 
magnetic flux ratio, it is often called 'the mass loss parameter' (Li 1995, 
Ferreira  1997),      
\begin{equation}\label{BPk}
\kappa = \sqrt{\frac{{\cal G}{\cal M}}{\varpi_{\star}V_{\star}^2}}
= 
- \sqrt{\frac{{\cal G}{\cal M}}{\varpi_{\star}V^2_{\rm p, A}}}   
\frac{\sin \theta_{\star}}{\cos \left(\psi_\star +\theta_\star \right) }  
= \kappa_{\rm BP}G^{-3/2}_o 
\,. 
\end{equation}
The constant $\lambda$ is the
specific angular momentum of the flow in units of $V_{\star}\varpi_{\star}$ 
and is related to the corresponding constant $\lambda_{\rm BP}$ in BP82,      
\begin{equation}\label{BPl}
\lambda = \frac{L}{V_{\star}\varpi_{\star}} =  
\lambda_{\rm BP} \kappa \sqrt{G_o}
\,. 
\end{equation}
The Bernoulli constant $\epsilon$ is the sum of the enthalpy, kinetic, 
gravitational and Poynting energy flux densities per unit of mass flux density  
divided by $V_{\star}^2$ (along $\alpha = 1$)
and is related to the corresponding constant 
$\epsilon_{\rm BP}$ in BP82,      
\begin{equation}\label{BPe}
\epsilon=\frac{E}{V_{\star}^2}=\epsilon_{\rm BP}\frac{\kappa^2}{G_o}
\,.
\end{equation}
Finally, the constant $\mu$ is proportional to the gas entropy, 
\begin{equation}\label{BPm}
\mu = \mu_{\rm BP}
\left[ 2 G_o^{3\gamma -4} \sin^{2\gamma - 2} \psi_o\kappa^{2\gamma}\right]   
\,. 
\end{equation}
In the above expressions, the label $o$ indicates the respective values of $G$ 
and $\psi$ at the base of the outflow. 
The correspondence between the parameter $\xi'_o$ in BP82 and our 
$\psi_o$ is 
\begin{equation}\label{BPxii}
\xi'_o=\cot \psi_o
\,.
\end{equation}

Note that in BP82 $\gamma$ does not appear since the outflow is cold and $\mu$, 
although it is defined, is never used. 
Also, a similar scaling exists for the parameters used in Li (1995) and 
Ferreira (1997), although with slightly different notations 
and a further relation between $x$ and $\kappa$ (cf. Eq. 28 in Ferreira 1997) 
due to the connection with a self-similar accretion disk thread by a large 
scale magnetic field.

\subsection{Numerical integration}

The numerical solution of Eqs. (A1)-(A3) requires the  fulfillment
of the regularity conditions at the positions of the three singularities 
(Alfv\'en and slow/fast modified magnetosonic critical points). 
This implies that the six parameters of the solution are not all independent. 
In the following 
we first shortly summarize the main properties of the critical conditions 
before we discuss the numerical procedure to obtain the solutions.

\subsubsection{Critical Points}          

It is evident that Eqs. (A2) and (A3) become indeterminate at  
the Alfv\'en surface where $G=M=1$. The regularity condition at this 
critical point can
be easily found together with the value of the derivative of $M^2$  
(see Appendix). 
Furthermore, the denominator of Eq. (A2) vanishes when the meridional 
component of the velocity $V_{\theta}$ satisfies the quartic (Vlahakis 1998):
\begin{equation}
\label{quartic}	
\displaystyle V_{\theta}^4-V_{\theta}^2 \left(C_s^2+V_A^2\right)+C_s^2 
V^2_{A\,,\theta} = 0
\,,
\end{equation}
where $C_s$ is the sound speed, and $V_A$ and $V_{A\,,\theta}$ the total and
meridional components of the Alfv\'en velocity, respectively. 

These singularities are typical `X-type' critical points, and
the above equation is the well known dispersion relation for the 
magnetosonic waves. However it is crucial to see that these singularities 
appear not when the flow speed, but instead where its meridional component 
coincides with the meridional component of the slow/fast magnetosonic velocity.
 
Bogovalov (1994, 1996) and Tsinganos et al. (1996) have emphasized that the 
singularities in MHD steady flows do not always coincide with the positions  
where the flow and the magnetosonic velocities coincide, but with the limiting 
characteristics, i.e., the FMSS and the SMSS.  
In our case the separatrix is found where $V_{\theta}$ is equal to either one of 
the triplet of the characteristic speeds ($V_{s, \theta}, V_{A,\theta}, 
V_{f,\theta}$). 
This is so because in addition to the azimuthal 
direction $\hat \phi$ due to the assumed axisymmetry, we have a second symmetry
direction, which is the radial direction $\hat r$ because of the assumed 
radial self-similarity. Therefore a compressible 
slow/fast MHD wave that preserves those two symmetries can only 
propagate along $\hat \theta$ which is perpendicular to both $\hat \phi$ 
and $\hat r$; the speed of propagation of such a wave satisfies exactly the 
quartic Eq. (\ref{quartic}) (for details see Tsinganos et al. 1996).

It is obvious that a physically acceptable solution with low velocity 
and high density at the base, but high speed and low density asymptotically 
must smoothly cross at least the SMSS and the Alfv\'en singularity.
Such  solutions have been widely analyzed in previous papers and are consistent
with the observational data on collimated stellar jets. However it is 
unescapable that also the fast magnetosonic singularity 
should be regularly crossed in order to have a
steady structure causally disconnected from the asymptotic region, where the
jet interacts with the environment (Bogovalov 1994).

\subsubsection{Numerical technique for the search of solutions}

An inherent difficulty of the problem is due to the fact that the positions 
of the previous critical points are not known a-priori, but need to be  
calculated simultaneously and selfconsistently with the sought for solution. 
At these critical points we do know some relations between various functions,  
for example, at the Alfv\'en surface the regularity condition, Eq. (A4),
should be satisfied. However, this knowledge alone is 
not practically enabling us to directly find a solution.
 
The way we will follow to construct a solution through all critical points 
is to use the  shooting method with successive iterations.
By starting the integration from an angle $\theta=\theta_i$ we reach  
a singular point  where, e.g.,  the denominator in
$dM^2/d \theta$ vanishes,  but not the numerator.
We then go back  and change some parameter and 
integrate again until it converges, i.e., the denominator \underline{and} the 
numerator vanish simultaneously.
A similar procedure is followed  to cross the other  singularity.
A rather key point is the choice of the starting position of integration.
Most of the previous studies solved the equations 
by starting  from the equator (BP82, CL94), or from infinity 
(Ostriker 1997).  It occurred to us that it is more convenient to
integrate the equations starting from the Alfv\'en critical point, i.e. from the 
conical surface $\theta=\theta_{\star}$, and move upstream (towards the
base) and downstream (towards the external asymptotic region). 

For the numerical integration, besides the parameters, we need also to 
choose the value of the colatitude  $\theta_{\star}$  and the value 
there of the slope of the square of the Alfv\'en number ($p_{\star}=
{{\rm d}M^2}/{{\rm d} \theta}|_{\theta_\star}$) together with the angle of 
expansion of the poloidal streamlines ($\psi_\star$).  
Some of these quantities must be tuned  to fulfill
the singularity conditions at the three critical points. It turned out
convenient for the assumed numerical technique to tune the values  
of $\lambda$ and $p_{\star}$ for getting the critical solution.

Hence, we first prescribe the parameters $\gamma$, $x$, $\lambda$ and $\kappa$, 
as well as $p_{\star}$, $\theta_{\star}$ and $\psi_{\star}$ while $\epsilon$ is 
deduced from the Bernoulli equation, Eq. (A3),
and $\mu$ from the regularity  condition at the Alfv\'en point, Eq. (A4).
The integration can now start upstream from $\theta=\theta_{\star}$  and the 
SMSS is encountered, but we cannot pass through it as, e.g., the denominator of 
$dM^2/d\theta$  vanishes there, but not the numerator.
We integrate again with different values of $p_{\star}$  until we find the 
opposite behaviour around the slow magnetosonic singularity  
(the numerator vanishes but not the denominator). Iteratively,
by fine tuning the value of $p_{\star}$, a solution is finally 
found which pass through the  SMSS.

Then we integrate downstream of the Alfv\'en surface and the  FMSS
is encountered, but in general it is not crossed.
Changing the value of the parameter $\lambda$ we integrate upstream 
again tuning to a new value of $p_{\star}$  until the  SMSS is crossed. Then we 
integrate downstream towards the fast 
magnetosonic singularity, and repeat all the procedure until we find the right 
values of $p_{\star}$ and $\lambda $ that allow the crossing of the two
singularities. At this point the complete solution is obtained by integrating,
with the correct values for all the parameters, upstream to the
base and downstream  towards the asymptotic region. 
 
\subsubsection{Selection of the parameters and boundary conditions}

In this study, the critical solution depends on the two `model' parameters 
($\gamma$, $x$) and the three independent `fieldline' parameters ($\kappa$, 
$\theta_{\star}$, $\psi_{\star}$).  The remaining ones ($\epsilon$, $\mu$,
$p_{\star}$, $\lambda$) are deduced from the Bernoulli equation and the 
crossing of the Alfv\'en, slow magnetosonic and fast magnetosonic 
singularities, respectively. 
This is consistent with the analysis of Bogovalov (1997) where it is argued 
that since the number of equations must be equal to the number of independent 
boundary conditions, a unique solution can be found if this number of 
independent 
boundary conditions equals to the number of outgoing waves generated at the 
reflection of a plane wave from that boundary. 
In $t$-dependent polytropic MHD there are 7 equations and 7 unknowns:  
the density, the pressure, the 3 components of the velocity and the 2 components 
of 
the magnetic field. There are also 7 waves: the entropy wave and the 
outwards/inwards propagating slow, Alfv\'en and fast MHD waves.   
So, we need 7 parameters with both counts, as expected. 
Now, if the boundary of the outflow is in the subslow region the number of {\it 
outgoing} waves from this boundary is 4, i.e., the entropy, slow, Alfv\'en and 
fast MHD outgoing waves. Subtracting the number of the boundary conditions we 
are left with 3 independent parameters, precisely  
$\kappa$, $\theta_{\star}$ and $\psi_{\star}$.  
Note that the polytropic index $\gamma$ and $x$ should not be included in this 
count since they are model parameters and do not depend on a particular 
streamline. 
%
\begin{figure*}
\centerline{
\psfig{figure=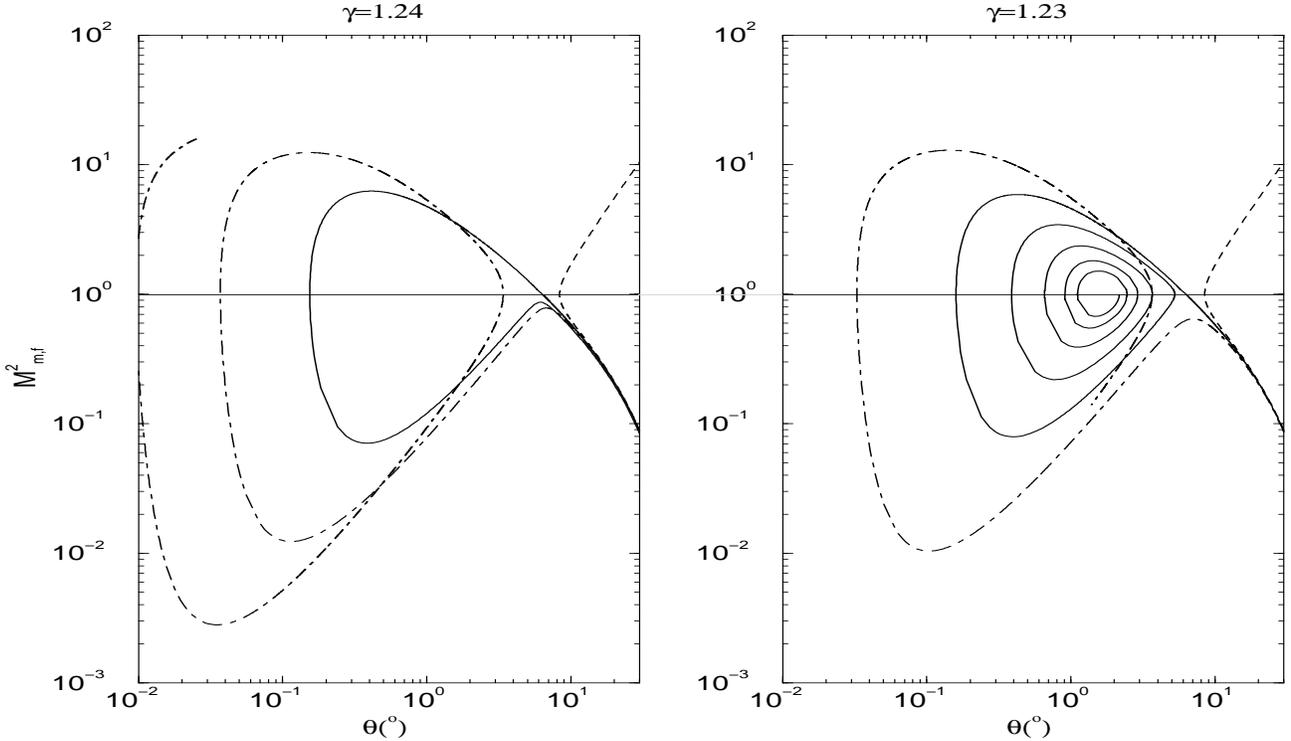,height=10.0truecm,width=17truecm,angle=270}
}
\caption
{Solution topologies around the  FMSS for 
$\gamma=1.24$ (left) and $\gamma=1.23$ (right). The other parameters are 
shown in Tab. 1. The critical solution drawn with a solid line crosses the 
line $M_{\rm m,f} = 1$ at 
$\theta = \theta_{\rm m,f}\approx  6^{\circ}$ while
the noncritical topologies  are plotted with dashed and dot-dashed lines.
For $\theta > \theta_{\rm m,f}$ and $M_{\rm m,f} < 1$ the three different 
branches of the solutions overlap.}
\end{figure*}

The integration is terminated in the upstream region when $M \rightarrow M_o<1$ 
and $G \rightarrow G_o < 1$, $\theta \rightarrow \theta_o\approx \pi /2$ 
and $\psi=\psi_o$. In all the 
calculations presented here we were able to follow the solution up to the 
equator, i.e. $\theta_o = \pi/2$. This base
should be in principle the disk surface, where our solutions should
consistently fit particular boundary conditions. 
Such an approach has been followed 
by Ferreira (1997) who looked for inflow/outflow MHD  solutions  with a 
consistent matching on the disk surface (see also Li 1995, 1996). 
This implies some further constraints on the parameters. For instance,
{\it if} the disk is also self-similar with a large scale magnetic field, a  
relation between  $\kappa$ and $x$ is expected, i.e.,  the mass loading in 
the outflow and the magnetic flux distribution on the disk. In addition, if
the outflow carries away all the angular momentum from the disk $x$ must be 
related to $\lambda$.  
 
As described above, the procedure to obtain a critical solution is  
extremely lengthy and rather time consuming. We must in fact
approach as close as possible the singularities ($\Delta \theta \approx
10^{-3}$), and this requires the determination of the parameters up to several 
digits. As we are mainly concerned to analyze the general behaviour of 
superfast 
magnetosonic solutions, in the present study we do not investigate the details
of the boundaries of the outflow. Therefore we assume that
between the base of the wind and the disk surface there is a thin `transition' 
region that allows the connection of the wind with the disk.

For similar reasons the present analysis has been performed only for a 
limited set of values of the parameters. We have fixed the `fieldline' 
parameters $\theta_{\star} = 59^{\circ}$, $\psi_{\star}
= 40^{\circ}$ and $\kappa = 2$, while two values have been assumed
for the `model' parameter $x$:  
0.75 (model I)  and 0.7525 (model II). In this two cases we will assume that  
the polytropic index $\gamma < 5/3$, i.e. some amount of heating occurs in 
the plasma. To make a comparison with a purely magnetocentrifugally driven 
outflow, we shortly discuss also the very general properties of an adiabatic 
solution ($\gamma=5/3$) assuming 
$\theta_{\star} = 60^{\circ}$, $\psi_{\star} = 45^{\circ}$, $\kappa = 3.873$
and $x=0.75$  
(i.e., values of $\theta_{\star}$ and $\psi_{\star}$ very close to those used
for the nonadiabatic models I and II on purpose have been selected). 

In the following two Sections we outline the main 
properties of the topologies around the FMSS
and discuss the structure of the critical solutions.

\section{Solution Topologies}

We present here the topology of two solutions around 
the fast magnetosonic point, assuming $\gamma=1.24$ and $\gamma=1.23$ for 
fixed $x=0.75$. The two slightly different values of the polytropic index define 
the transition between two families of topologies. This drastic change in the 
topological behavior of the solutions in the neighborhood  
of the X-type point illustrates the 
difficulty of exact crossing the fast magnetosonic point. The  parameters 
for the various cases are listed in Tab. 1, 
while in Fig. 1 we plot the two sets of topologies for the superfast 
magnetosonic number $M_{\rm m,f} (\theta )$. Note that this plot is 
obtained from a projection of the solutions from the 3-D space of 
$M(\theta)$, $G(\theta)$ and $\theta$ to the plane $M_{\rm m,f}$ -- $\theta$.
This three-dimensional structure of the topology explains why some of the lines 
obtained by projection are crossing each other (see for another such example
Tsinganos \& Sauty 1992). This feature of course does not appear in more  
classical topologies of one-dimensional solutions, e.g., Weber \& Davis (1967). 
 
In the first case ($\gamma=1.24$)
three solutions are plotted in Fig. 1a for different values
of $\lambda$ and $\mu$. The {\it critical} solution (solid line in Fig. 1a),
moving downstream in the direction of decreasing $\theta$ crosses the FMSS at 
$\theta_{\rm m,f} \approx 6^{\circ}$, has a maximum at $\theta \approx 
0.4^{\circ}$
and then at $\theta \approx 0.15^{\circ}$ crosses back the $M_{\rm m,f}=1$ 
line but with an infinite slope moving towards increasing $\theta$. 
Then, this solution continues marching towards increasing $\theta$ 
and remains always subfast magnetosonic, with $M_{\rm m,f}$ reaching a 
maximum at $\theta \approx \theta_{\rm m,f}$. 

By slightly decreasing $\lambda$ the solution crosses the $M_{\rm m,f}=1$ 
line with infinite slope at $\theta > \theta_{\rm m,f}$  (dashed line in 
Fig. 1a). Conversely, for a slightly larger value of $\lambda$ the solution 
(dot-dashed line in Fig. 1a) reaches a maximum at $\theta \approx 
\theta_{\rm m,f}$ remaining subfast magnetosonic 
(i.e. it behaves like a `breeze' solution) and becomes superfast 
magnetosonic with diverging slope at $\theta \approx 0.04^{\circ}$. 
Then, this solution remains always in the region 
$\theta < \theta_{\rm m,f}$, with a spiraling behaviour, i.e., by crossing  
many times up and down the $M_{\rm m,f}=1$ transition with infinite slope. 
Note that the solution shown in Fig. 10 of Ferreira (1997) probably belongs to 
this family of non critical solutions. 

For $\gamma = 1.23$ (Fig. 1b) the topology of the non critical solutions  
remains the same. The critical solution however shows a different behaviour 
remaining always in the region $\theta < \theta_{\rm m,f}$ by spiraling
around the $M_{\rm m, f}=1$ transition (solid line in Fig. 1b). 
 
The topological structure of our solutions implies that downstream of the
FMSS a focal critical point must be present, so that no
solution can asymptotically reach $\theta =0$ with superfast magnetosonic  
speeds. This ought to be expected from the construction of this 
model where we should have  
$\lim_{\theta \to 0} {V_{\theta}}/{V_{{\rm f},\theta}}=0$,  
if we have a cylindrically collimated outflow. In other words, 
the surface $M_{\rm m, f}=1$ needs to be crossed again with a downstream 
superfast/subfast magnetosonic transition (see also Contopoulos 1995).  
At the same time, we should keep in mind that these radially self-similar 
solutions are not valid to model outflows around the rotational axis, because 
of their singular behaviour there.

\begin{table}
{\caption{Parameters of the topological solutions$^{\rm a}$}}
\begin{center}
\begin{tabular}{|c|c|c|c|c|c|c|c|c|c|c|}
\hline
$\gamma$ & $p_{\star}$ & $\lambda^2$ &  $\mu$  &  Solution\\
\hline
1.24  &   -12.5522 & 72.7220  & 6.7825  & critical (solid)\\
      &            & 72.0000  & 6.9069  & terminated (dashed)\\ 
      &            & 73.0000  & 6.7347  & spiral (dot-dashed)\\
\hline
1.23  &   -12.6468 & 75.8919  & 6.6983  &  critical (solid)\\ 
      &            & 75.0000  & 6.8506  &  terminated (dashed) \\ 
      &            & 77.0000  & 6.5092  &  spiral (dot-dashed)\\ 
\hline
\end{tabular} 
\vskip 0.1 true cm
 $^{\rm a}$  assuming $x=0.75$, $\theta_{\star}=59^{\circ}$, 
 $\psi_{\star}=40^{\circ}$ and $\kappa = 2$.
\label{tab:table1}
\end{center}
\end{table}
\vspace{0.5cm}
Note that not all solutions with $M_{\rm m,f} > 1$ are physically acceptable 
because they become subfast magnetosonic, crossing the singularity with 
diverging slope and therefore they are multivalued for the same $\theta$.   
Hence, these solutions could correspond  
to the terminated solutions in Parker's terminology for the solar wind 
with one (Parker 1958), or,  multiple critical points (Habbal \& Tsinganos 
1983). Nevertheless,  the present critical
solutions are causally disconnected from the inner region of the flow, so that 
they could be stopped by suitable boundary conditions, e.g. through a shock 
with the external medium at some angle $\theta_{min} < \theta_{\rm m,f}$ 
without affecting the structure of the outflow upstream of the FMSS.

It is worth to mention that, from the technical point of view, the main 
difficulty in obtaining a critical solution is the fact that all solutions 
(critical ones as  well 
as non critical ones) always reach $M_{\rm m,f} = 1$ with infinite slope at some 
angle $\theta$. They
become ``terminated'' at this point, even if they belong to the family 
of the dot-dashed solution family of Figs. 1. And, both families of non critical
solutions almost coincide far from the vicinity of the critical X-point. 
This is the reason why the crossing of the critical point is so difficult.

\section{Results}

The values of the parameters in the previous Section were chosen such as 
to illustrate the topology of the solution around the fast critical point. 
However, they do not correspond to some interesting critical solution 
from the astrophysical point of view. 
For example, the fast magnetosonic transition occurs for a rather slow 
velocity and not far from the Alfv\'en critical surface. We found that 
much more interesting results are obtained for a flow closer to isothermal
conditions. We then discuss in the following the properties of solutions 
obtained with $\gamma=1.05$ and for two sets of the remaining parameters 
(models I and II in Tab. 2).

\begin{table}
{\caption{Parameters of the solutions$^{\rm b}$}}
\begin{center}
\begin{tabular}{|c|c|c|c|c|c|}
\hline
model & $x$ & $p_{\star}$ & $\lambda^2$ & $\mu$  &  $\epsilon$  
\\
\hline
I & 0.75    &   -14.07 & 136.9232 & 2.9902    &  156.617    \\
\hline
II & 0.7525  &   -14.02 & 136.2261 & 3.1715    &  158.233    \\
\hline
\end{tabular} 
\vskip 0.1 true cm
 $^{\rm b}$assuming $\gamma=1.05$, $\theta_{\star}=59^{\circ}$, 
 $\psi_{\star}=40^{\circ}$ and $\kappa = 2 $. 
\label{tab:table2}
\end{center}  
\end{table}

In both cases $\theta_{\star}=59^{\circ}$, 
$\psi_{\star}=40^{\circ}$ and $\kappa = 2$, as in the previous topological 
analysis, with $x=0.75$  and $x=0.7525$. 
The remaining parameters are deduced from the requirement to fulfill the 
criticality conditions and are listed in Tab. 2.
We remark that this different choice on the scaling of the magnetic field $x$
is important to connect the solution to an accretion disk in the spirit of 
what has been  done by Li (1996) and Ferreira (1997). 
In such a case, a value of $x$ larger than some minimum above the
value of BP82, $x=0.75$, is necessary to allow ejection 
(Ferreira \& Pelletier 1995). However, it does not mean that our 
solution fulfills all requirements to connect to such a disk, as we discuss
later. The main goal here is to show that the solution is not affected 
qualitatively by the change in $x$ as far as the crossing of critical points
 is  concerned.

\setbox1=\vbox{\hsize=10truecm\vsize=10truecm
\psfig{figure=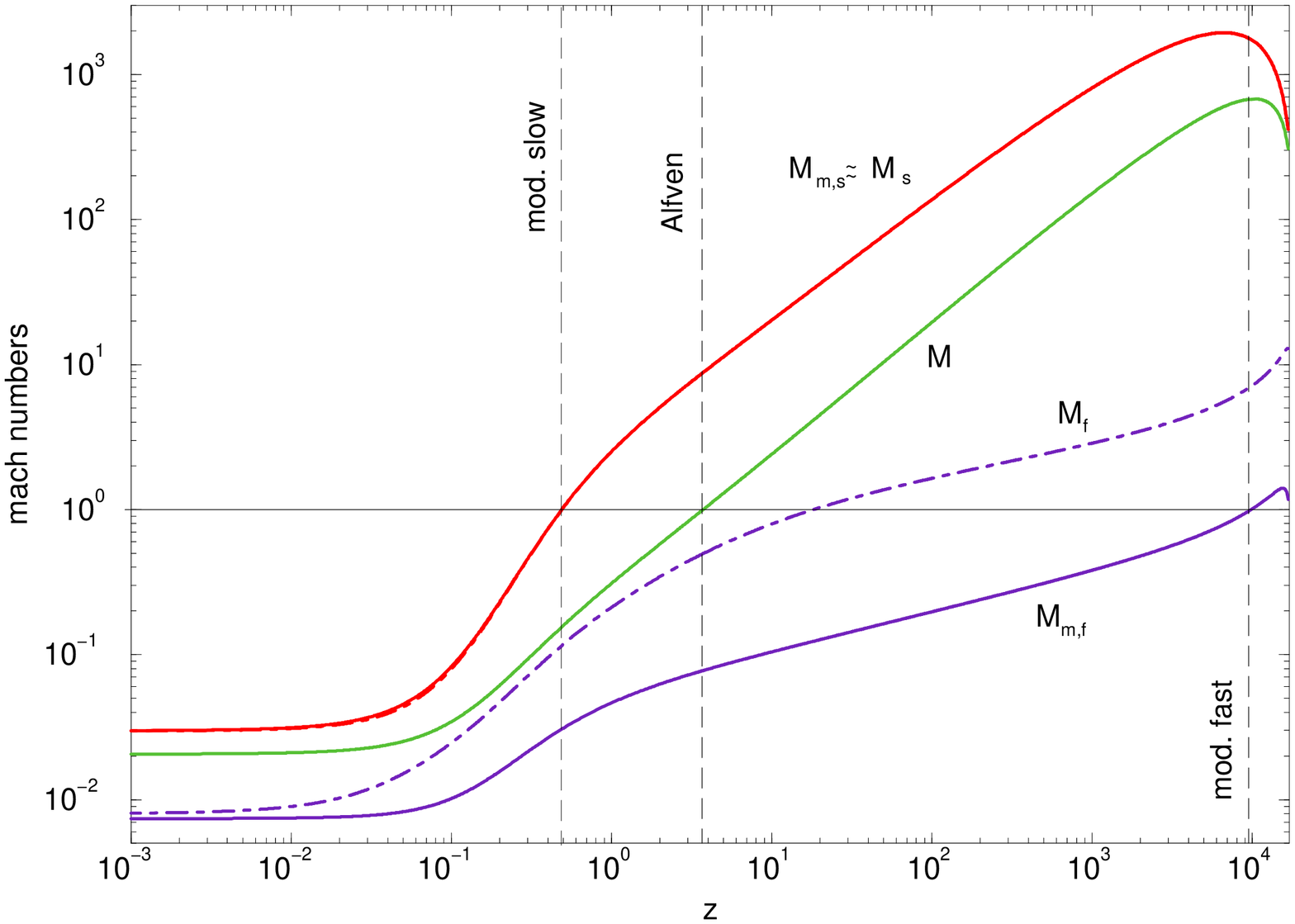,height=10.0truecm,width=9truecm,angle=270}}
\par
\begin{figure} 
\centerline{\box1}
\vspace{-1cm}
\caption
[The various Mach numbers for the polytropic $r-$ self-similar model]
{Plot of the various Mach numbers {\it vs.} the vertical height $z$ in 
units of the equatorial cylindrical radius $\varpi_o = \varpi (z=0)$ of a 
particular 
fieldline. A polytropic radially self-similar model is used with 
the parameters of model I (Tab. 2). The critical transitions at the three 
singularities are marked with 
vertical lines.}
\label{mach35}
\end{figure}

\setbox1=\vbox{\hsize=9truecm\vsize=9truecm   
\psfig{figure=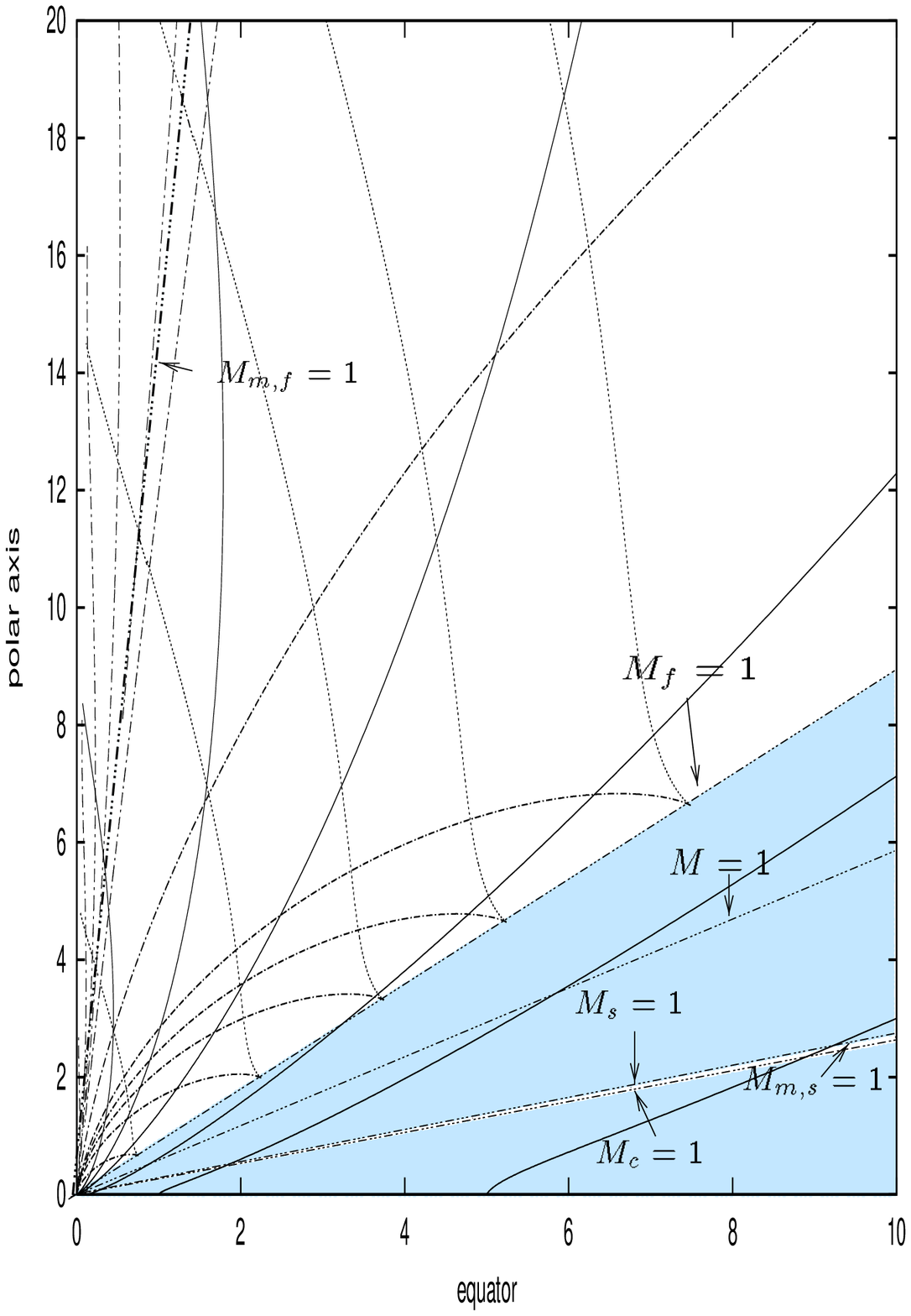,height=16.0truecm,width=16.0truecm,angle=180}}  
\par
\begin{figure*}
\centerline{\box1}
\caption
[The fieldlines and the characteristics for the polytropic $r$- self-similar 
model]
{Poloidal fieldlines (solid), characteristics (dotted and dot-dashed)
and cones of the singular surfaces (dot-dot-dashed)
are shown for the radially self-similar polytropic solution of model I.
In the shadowed regions the governing partial differential equations are of
elliptic type and no characteristics exist. The parameters are as in Fig. 2.
}
\label{linescharact35}
\end{figure*}

In Fig. 2 we plot the various Mach numbers along each field 
line $\alpha$ {\it vs.} the vertical height $z$ in 
units of the equatorial cylindrical radius $\varpi (z=0)$ of a particular 
fieldline and for model I. The various critical transitions are indicated, 
and on the disk surface we find $G_o \approx 0.16$ and $M_o \approx 
0.02$. The SMSS almost coincides with the point where 
the flow becomes superslow magnetosonic 
($ M_{\rm s, m} \approx M_{\rm s} =1$), at $z \approx 0.5$. 
The Alfv\'en critical point ($M=1$) is crossed at 
$z \approx 3.5$ while the wind becomes superfast magnetosonic ($M_{\rm f}=1$) 
at $z \approx 20$. Much farther away is the  
FMSS, at $z \approx 10^4$. Downstream of this position
the various Alfv\'en numbers decrease, as expected from the previous 
topological analysis. 

The turning of the solutions is evident in Fig. 3, where the poloidal
streamlines together with the characteristics are plotted. 
They cross all critical surfaces, and for
$\theta < \theta_{\rm m,f}$ the fieldlines converge towards the symmetry
axis such that the conical region with $\theta < \theta_{\rm m,f}$ is causally 
disconnected from the rest of the domain. The two families of the 
characteristics in the hyperbolic domain bounded by the cusp and slow surfaces 
are better seen in Fig. 4 obtained for 
the adiabatic case, with 
a different set of 
parameters. One family of characteristics (black) is tangent to the SMSS at 
$M_{\rm m, s}=1$ while the other (grey) crosses it. 
Similarly, in the hyperbolic domain 
bounded by the cone where $M_{\rm f}=1$ one of the family of the characteristics 
(black) is tangent to the FMSS indicated by $M_{\rm m, f}=1$ while the other 
(grey) crosses it.  We remind that the cusp surface ($M_{\rm c}=1$) does not 
coincide with any singularity or typical velocity in the flow.

The components of the outflow speed along a line $\alpha=const$ in 
units of the initial $z$-component of the flow speed at the disk, are 
plotted in Fig. 5.
The units are choosen in order to make a direct comparison of this solution with
other solutions in the literature (e.g. BP82).
Close to the disk level, the escape speed is high, $V_{\rm esc, o} \approx 440$, 
the initial rotational speed is lower, $V_{\rm \phi , o} = 110$ and of the 
order of the Keplerian speed, $V_{\rm Kep} \sim 3 V_{\rm \phi , o}$. 
The azimuthal speed $V_{\phi}$ after some increase in the region of 
corotation, approximately up to the Alfv\'en critical 
point at $z\sim 4$, decays to zero transferring its corresponding 
kinetic energy to poloidal motion. Thus, the $z$- and $\varpi$-components 
of the poloidal motion grow from their subslow and subescape values at 
the disk level where $V_z = 1$ to the high values obtained at 
the modified fast critical point where $V_z \sim 10^3$. The poloidal 
speed exceeds the local escape speed around the Alfv\'en transition.        
A comparison of model I and II makes clear that the
different values of $x$  do not strongly affect the global behaviour of the 
solutions, even though the boundary conditions of the disk are rather 
different.   
\setbox1=\vbox{\hsize=9truecm\vsize=9truecm  
\psfig{figure=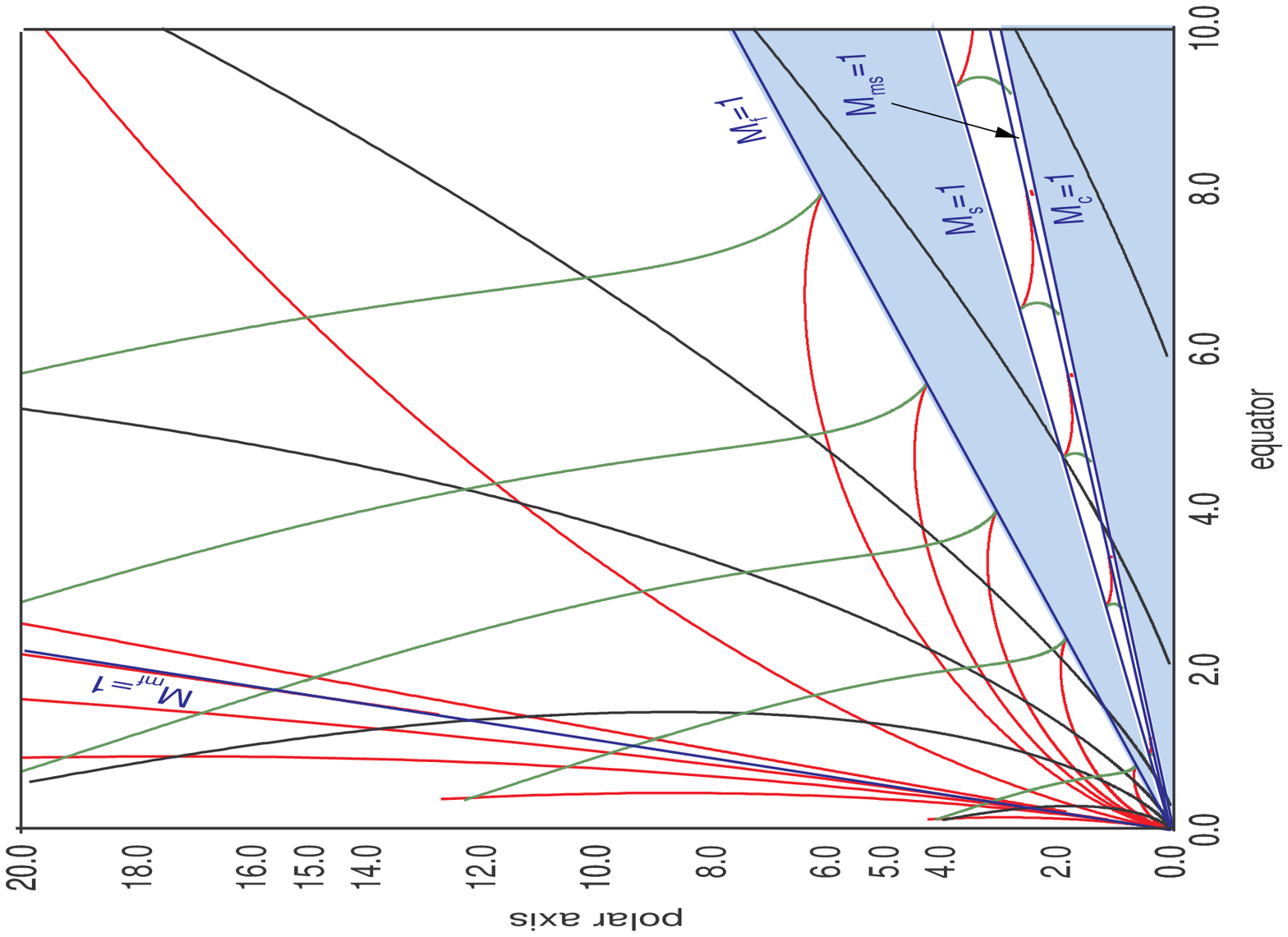,height=16.0truecm,width=16truecm,angle=270}}
\par
\begin{figure*}
\centerline{\box1}
\caption
[The fieldlines and the characteristics for the polytropic $r$- self-similar 
model]
{Poloidal fieldlines (solid) and characteristics (gray, black)
for a radially self-similar polytropic (adiabatic) model with $x=0.75$, 
$\gamma = 5/3$, 
$\theta_{\star}=60^{\circ}$, $\psi_{\star}=45^{\circ}$, $\kappa^2 = 15 $,
$\mu=10.9239$, $\lambda^2=2.7935$, $p_\star=-5.5744$, $\epsilon=9.4487$. 
The cones of the different transitions are shown, as well as the regions where
equations are elliptical (shaded).}
\label{alll}
\end{figure*}

\setbox1=\vbox{\hsize=9truecm\vsize=9truecm  
\psfig{figure=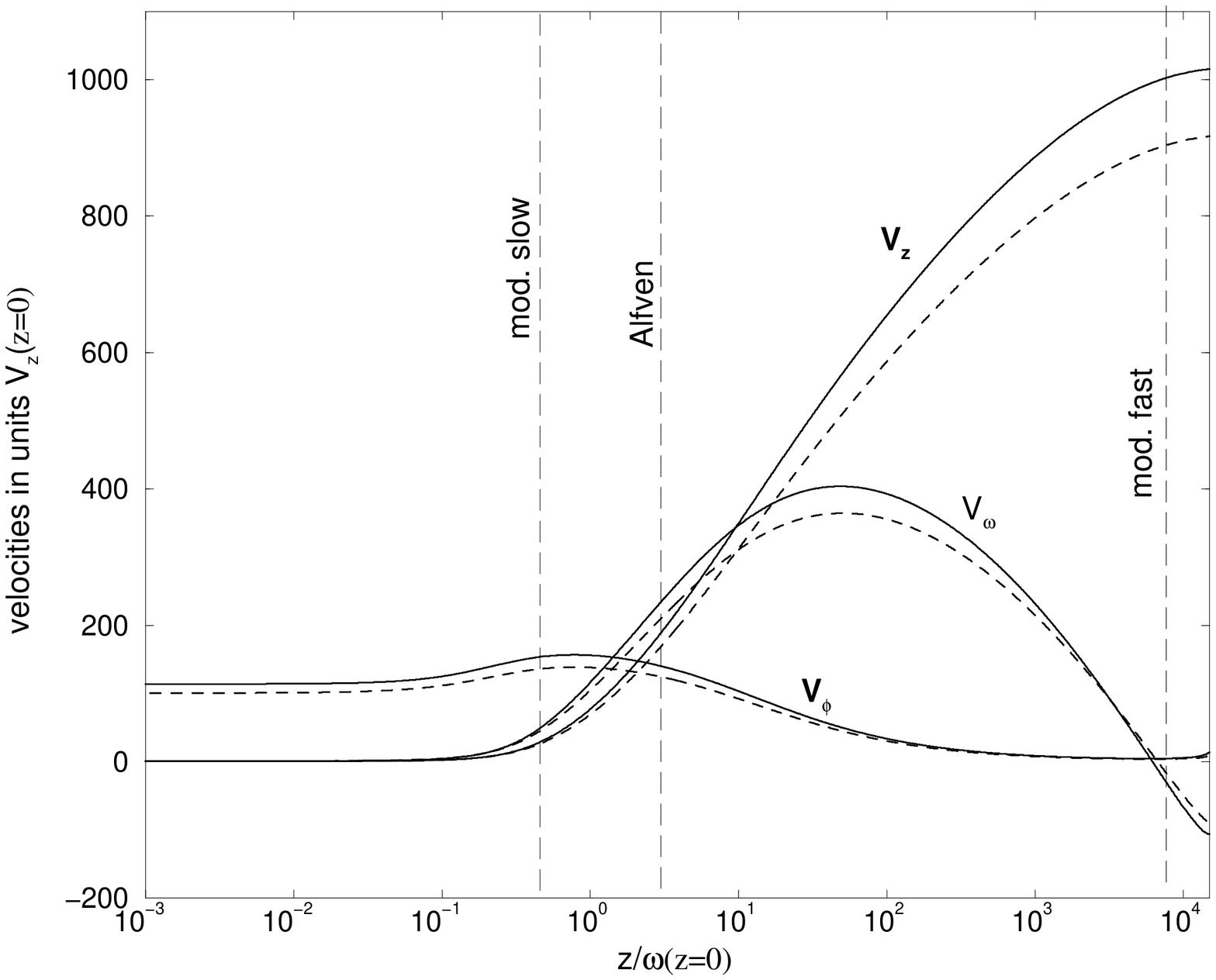,height=8.0truecm,width=8truecm,angle=270}}
\par
\begin{figure}
\centerline{\box1}
\caption
[Velocities-pressures]
{Plot {\it vs} $z$ of the components of the outflow speed   for model I (solid) 
and model II (dashed), with the parameters given in Tab. 2.
}
\label{vel3547new}
\end{figure}   

Downstream of the Alfv\'en transition the azimuthal component of the magnetic 
field grows to very high values in comparison to the poloidal component 
(Fig. 6).  At the modified fast critical point practically all the magnetic flux 
is in the azimuthal direction. For example, $B_{\phi}/B_P \approx 1$ at the 
disk, while $B_{\phi}/B_P \approx 60$ after the modified fast transition for 
both, models I and II. 
 From Fig. 6 it may be also seen that 
the flow velocity is largely in the $z$-direction with very small 
components along $\hat \phi$ and $\hat \varpi$.   
In Fig. 6, the main difference between the two models is in the
region upstream of the SMSS: for $x=0.75$ the angle between the poloidal 
fieldline and the disk surface is $\psi_o \approx 67^{\circ}$, while for
$x=0.7525$ this angle is $\psi_o \approx 56^{\circ}$. 
Although these values are not very 
different, only the second case matches the outflow launching condition for a 
cold plasma given in BP82. This means that magnetocentrifugal driving is more 
efficient in model II at the base. However, we note that at the SSMS where the 
plasma pressure has dropped significantly both solutions can be 
magneto-centrifugally accelerated. The end result shown in Fig. 5 is that the
terminal speed is lower when $x$ is larger, i.e., when the ejection index is 
higher. This  result is consistent with Ferreira's (1997) analysis.

The behaviour of the various components of the conserved total energy 
$E$ {\it vs. }  
$z$, plotted in Fig. 7, provides information on the different driving mechanisms 
that govern the dynamics of the outflow. 
Upstream of the SMSS and close to the base, most of the energy flux is  
electromagnetic plus some amount of enthalpy. The kinetic energy of the plasma 
is  negligible. As the slow magnetosonic surface  
is approached, the kinetic energy sharply increases with a corresponding  
decrease of the thermal energy. Downstream of the Alfv\'en surface  the  
Poynting flux rapidly decreases;  
the poloidal kinetic energy keeps increasing, becoming largely the
main component of the energy flux at the position of the FMSS. This behaviour is
basically the same for both models I and II. 
In order words, there is some contribution to the acceleration of thermal origin 
up to the modified slow critical point after which the pressure drops to a 
rather 
constant  value while the magnetic pressure maintains considerably higher values 
up to the Alfv\'en transition. 
 
\begin{figure}
\centerline{
\psfig{figure=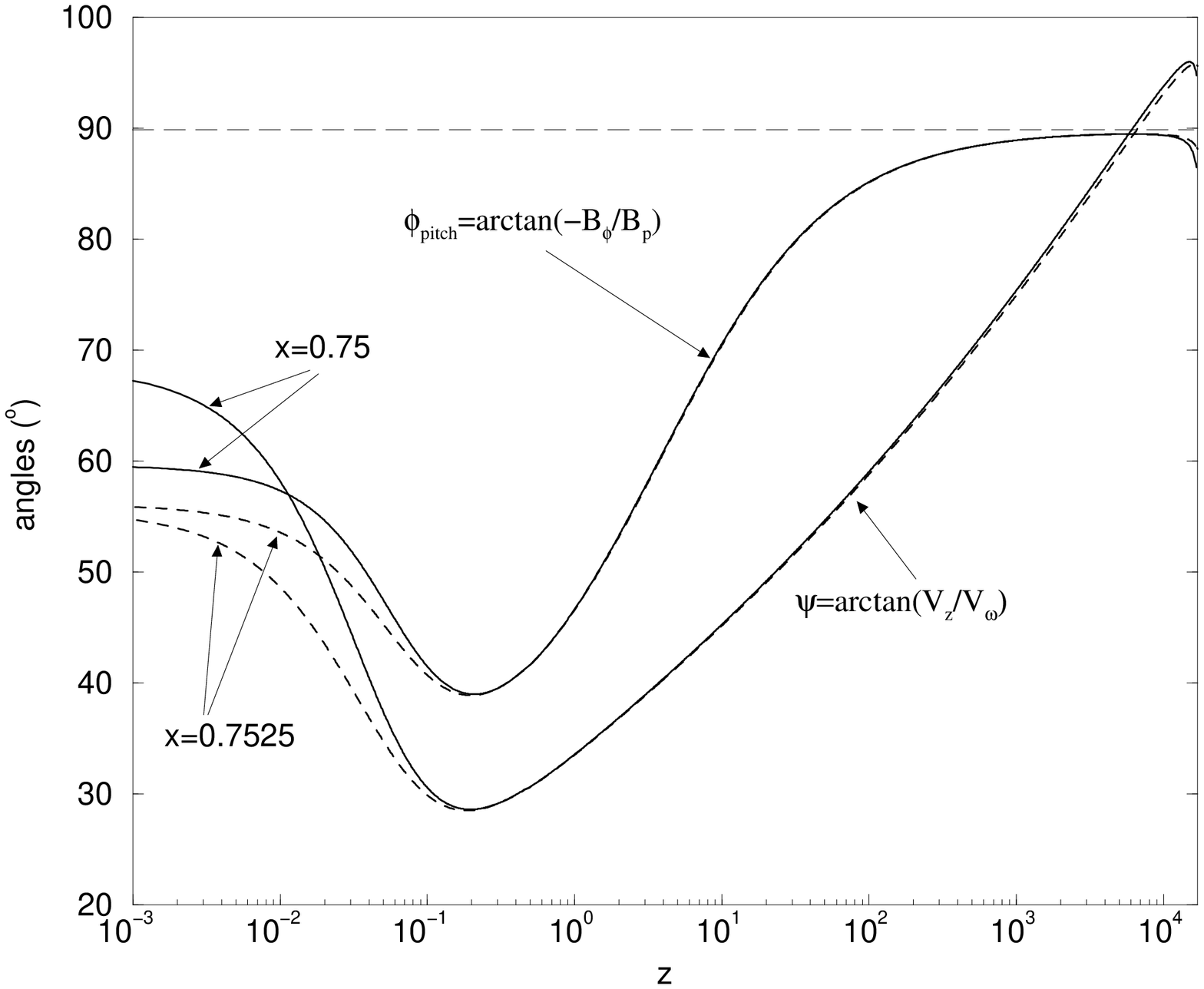,height=8.0truecm,width=8.0truecm,angle=270}} 
\caption 
{Plot {\it vs.} $z$ of the ratios of the magnetic components $B_{\phi}/B_p$ 
and poloidal flow speeds $V_z/V_{\varpi}$ for models I and II (solid and 
dashed lines), with the parameters given in Table 2.}
\end{figure}

\begin{figure}
\centerline{
\psfig{figure=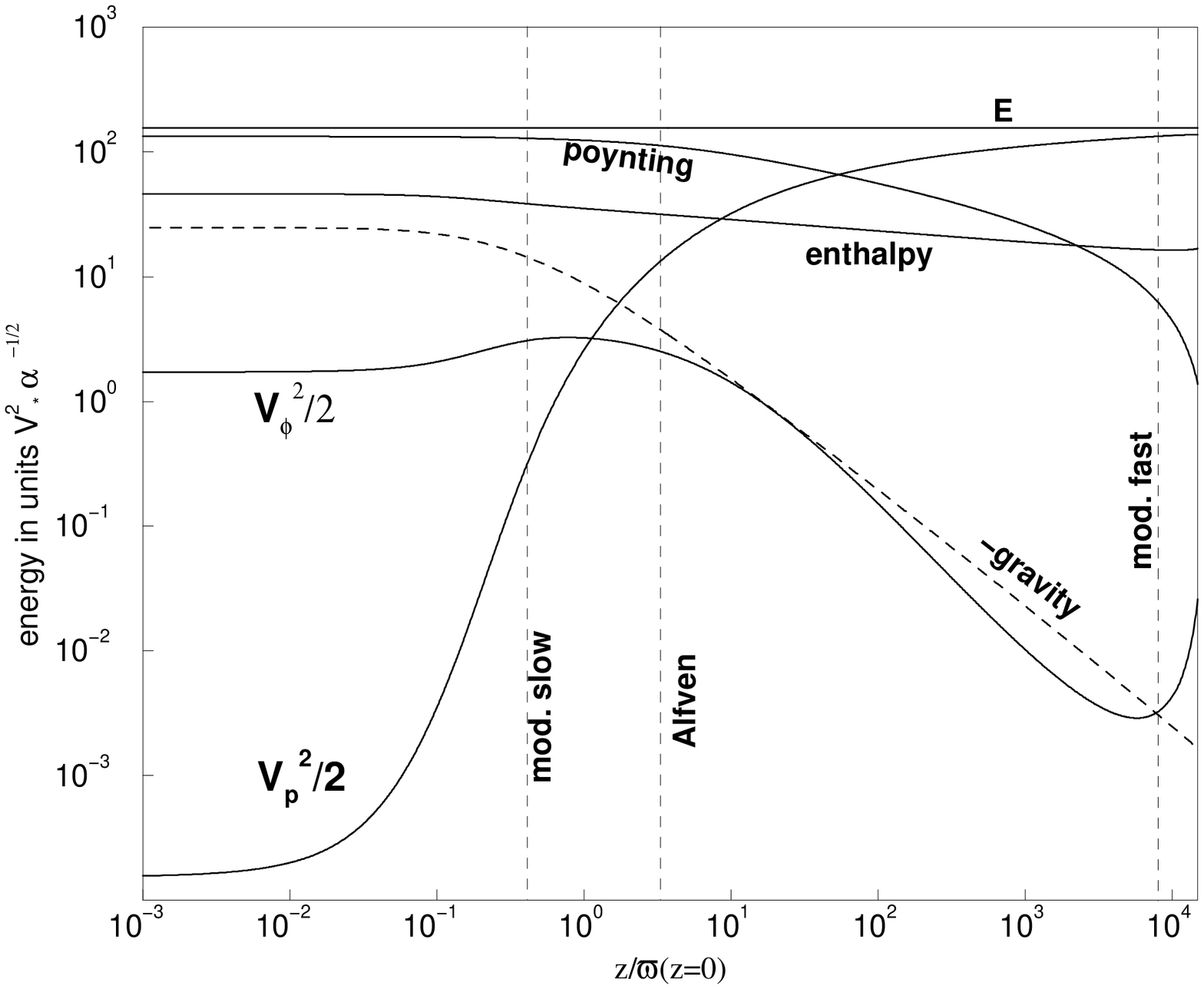,height=8.0truecm,width=8truecm,angle=270}} 
\caption 
{Plot {\it vs.} $z$ of the various components of the conserved total energy $E$ 
for model I.}
\end{figure}

We conclude this section by pointing out that the 
two solutions we have analyzed here correspond to efficient 
magnetic rotators in the terminology of Bogovalov \& Tsinganos (1999),  
since the ratio of the corotational velocity to the poloidal Alfv\'en velocity 
at the Alfv\'en critical surface (the parameter $\alpha$ in their notation)
has a value greater than unity ($\approx 2.13$).

\section{Discussion}

Before discussing the main physical implications of our results, also in 
connection with those obtained by other authors, we show that the present
solutions are suitable to describe the physical properties of astrophysical
outflows.
 
\subsection{Astrophysical applications}

The modeling of a particular astrophysical outflow requires first the 
calculation of all physical quantities from the non dimensional 
parameters characterizing the particular model. 
We will address here this question of calculating some observable quantities 
of disk-winds associated with protostellar objects from the parameters 
of our model. 

We deduce first the ratios of some
characteristic speeds at the disk level, keeping in mind that from the numerical
results we have obtained  $M_o \sim 0.01$ and $G_o \sim 0.1$. We will refer
in the following mainly to the solutions with $x=0.75$. 

{\it First}, the ratio of the poloidal Alfv\'en and Keplerian speeds at the disk 
level is:
\begin{equation}\label{rho_o}
\left( \frac{V_{Ap}}{V_{\rm Kep}} \right)_o = 0.316 \times
\left( \frac{M_o}{0.01} \right) \left(\frac{G_o}{0.1}\right)^{-3/2} 
\frac{1}{\kappa \sin\psi_o} \approx 0.178
\,.
\end{equation}
The poloidal magnetic field which is essential in the launching of the outflow 
is 
anchored in the disk and its energy density is less than the rotational 
kinetic energy density of the disk. Thus, the field is rather weak to brake the 
rotation of the plasma at the disk and it is carried passively around by 
azimuthal rotation.  
 
{\it Second}, the ratio of the sound and initial speeds at the disk level is:
\begin{equation}\label{CsVo}
\left(\frac{C_s}{V_o}\right)_o = 70.7 \times \left({\gamma 
\mu}\right)^{1/2}
\left(\frac{G_o}{0.1}\right)^{2}\frac{10^{-4}}{M_o^{(\gamma+1)}}
\approx 94  
\,,
\end{equation}
where $V_o=V_z (z=0)$. 
The initial ejection speed is negligible in 
comparison to the thermal speed at the disk, a situation similar to a 
thermally driven wind.  
 
{\it Next}, the ratio of the sound and Keplerian speeds at the disk level is:
\begin{equation}\label{CsVk}
\left(\frac{C_s}{V_{\rm Kep}}\right)_o = 0.22 \times (\gamma 
\mu)^{1/2}\left(\frac{G_o}{0.1}\right)^{1/2}\frac{1}{ \kappa M_o^{(\gamma
-1)}} \approx 0.314
\,. 
\end{equation}
We notice that the Keplerian speed is about 3 times 
higher than the thermal speed at the disk. Thus, thermal effects cannot 
inhibit the rotation of the disk.  
 
{\it Finally}, the ratio of the Keplerian and initial speeds at the disk level is:
\begin{equation}\label{VkVo}
\left(\frac{V_{\rm Kep}}{V_o}\right)_o = 316 \times \kappa  
\left(\frac{G_o}{0.1}\right)^{3/2}\left(\frac{10^{-2}}{M_o}\right)^2 
\approx 300  
\,,
\end{equation}
i.e., the initial speed is negligible in comparison to the Keplerian speed. 
 
In our case the flow speed at the fast critical point is 
about $10^3$ the initial speed $V_o$. In agreement with the observations we can 
reasonably assume  the terminal speed of the outflow to be 
$\sim 400$ km s$^{-1}$, 
such that  its velocity at the base is  $V_o = $ 0.4 km s$^{-1}$. 

In principle, radially self-similar models do not have an intrinsic scale 
length; however from the previous estimate of the initial speed one allows to 
calculate the footpoint of the  reference fieldline $\alpha=1$ on the disk. In 
units of 10 solar radii this cylindrical distance $\varpi_o$ is:
\begin{equation}\label{varpi_o}
\frac{\varpi_o}{10 R_{\odot}} = 0.19 \times  
\left(\frac{M_o}{0.01}\right)^4\left(\frac{0.1}{G_{o}}\right)^3
\frac{1}{\kappa^2}\frac{\cal M}{\cal M_{\odot}}
\left(\frac{V_o}{{\rm km} \, {\rm s}^{-1}}\right)^{-2}.
\end{equation}
Hence, for a one solar mass star we get $\varpi_{o} \approx  12.5 R_{\odot}$.

It is also interesting to calculate the mass-loss rate $\dot {\cal M}_w$ in 
units 
of $10^{-8} {\cal M}_{\odot} \, {\rm yr}^{-1}$ :
\begin{eqnarray}\label{mdot} 
\begin{array}{l}
\displaystyle \frac{\dot {\cal M}_w}{10^{-8}{\cal M_{\odot}} \,{\rm yr}^{-1}} = 
0.0386 \times  
\left ( \displaystyle \frac{M_o}{0.01}\right)^2
\left(\frac{B_{z,o}}{10 G}\right)^2 \times
\\ \\ 
\left( \displaystyle \frac{\varpi_o}{10 R_{\odot}}\right)^{2}
\left ( \displaystyle \frac{V_o}{\rm km \,{\rm s}^{-1} }\right)^{-1} 
f(\alpha_{out}, \alpha_{in} )
\,,
\end{array} 
\end{eqnarray}
where
\begin{equation}\label{mdot1}
f(\alpha_{out}, \alpha_{in} ) = 
\frac{\alpha_{out}^{x-3/4}-\alpha_{in}^{x-3/4}}{x-3/4}
\qquad {\rm if} \quad x\neq 0.75
\,,
\end{equation}
and 
\begin{equation}\label{mdot2}
f(\alpha_{out}, \alpha_{in} ) = 
\ln \frac{\alpha_{out}}{\alpha_{in}}  
\qquad {\rm if} \quad x= 0.75
\,.
\end{equation}
By assuming $\varpi_{in} = \varpi_{o}$, 
$\varpi_{out} \approx 10 \varpi_{o}$ and $B_{z,o}= 8$ G we have 
${\dot {\cal M}_w}/(10^{-8} {\cal M_{\odot}} \, {\rm yr}^{-1}) \approx 1$,  
with a temperature at  the  base of the flow of:
\begin{equation}\label{T_o} 
T_{o,in} = 3 \times 10^5 
\mu\left(\frac{G_o}{0.1}\right)^4\frac{10^{-8}}
{M_o^{2(\gamma+1)}} \left( \frac{V_o}{{\rm km}\, {\rm s}^{-1}} \right)^2  
\approx 8\times 10^4 \,{\rm K}
\,,
\end{equation}
\begin{equation}\label{T_o,out} 
T_{o,out} =
T_{o,in} \frac{\varpi_{in}}{\varpi_{out}}
\approx 8\times 10^3 \,{\rm K}
\,.
\end{equation}
We remind that $T_o$
is not the temperature of the disk as we have 
assumed a transition layer between the disk surface and the base 
of the flow (see Sec. 2). This region could be reasonably related to a corona 
heated by dissipative processes in the plasma (e.g. magnetic reconnection,
ohmic heating, etc.; see e.g. K\"onigl \& Pudritz 2000).
 
As the flow corotates roughly up to the Alfv\'en point (Fig. 5) the specific 
angular momentum carried by the wind is $\dot J_w = \dot {\cal M}_w \Omega 
\varpi_\alpha^2$ 
while the angular momentum that has to be extracted locally at the foot point 
$\varpi_o$ of the fieldline in order that the disk accretes at a rate 
$\dot {\cal M}_a$ is $\dot J_a= (1/2) \Omega \varpi_o^2 \dot {\cal M}_a$ 
(Spruit 1996). 
If the angular momentum carried by the wind is a fraction $f$ of $\dot J_a $ 
while $1-f$ is the fraction carried away by viscous stresses, then the 
ratio of the mass fluxes in the wind and in the accretion flow is
$$ 
{\dot {\cal M}_w \over \dot {\cal M}_a} = {f\over 2} {\varpi_o^2 \over 
\varpi_\alpha^2} \lesssim 0.015
\,,
$$ 
taking into account that $ \varpi_\alpha = 5.8 \varpi_o$ for model I. 
 It follows that the rate of the outflowing mass is at most 
of the order of 1$\%$ of the rate of the accreted mass; and this is achieved 
when the wind carries all the angular momentum of the accreted mass. When the 
outflow carries a smaller fraction of the angular momentum of the disk, the mass 
loss rate in the wind is an even smaller fraction of the mass loss rate in the 
wind.  
In other words, the mass loss rate in the wind is a negligible fraction of the 
accreted mass, despite that the jet may carry most of the angular momentum of 
the accreted mass. Similar results are obtained for the case $x=0.7525$. 
Therefore, from the above arguments we may conclude 
that from our solutions  we  deduce for the physical parameters values in 
reasonable agreement with those observed in this class of objects.

Our solution terminates at $z/ \varpi_o \approx 2 \times 10^4$, i.e., 
at $\approx 400$ Astronomical Units (AU) from the central star. At this
position we could argue that there exists a shock matching the solution with the
outermost region of the outflow (Gomez de Castro \& Pudritz 1993). It is well 
known that bright knots are observed 
on scales of thousands  AU along most protostellar jets. These 
configurations are shocks that are interpreted as originated either by fluid 
instabilities on the jet surface or by temporal variations in the velocity of 
the outflow (Burrows et al. 1996, Ray 1996, 
1998,  Micono et al. 1998, K\"onigl \& Pudritz 2000). It could be reasonable to 
associate the terminal shock of our solutions  with the inner knots, found at 
distances down to $\approx 100$  AU from the star. However we 
cannot ignore that these knots are non steady configurations and move  
outwards  with velocities $\sim 100 \div 200$ km s$^{-1}$ (Ray 1996). We could assume that 
the shock is
well upstream of the optical knots: polarimetric radio data on the T Tauri  
object are consistent with the presence of a shock at $\approx 20 \div 40$
AU from the star (Ray et al. 1997). Alternatively the terminal shock could 
indeed
be located approximately at the positions of the inner knots, but there the flow
looses both self-similarity and steadiness. However as we are in the superfast
magnetosonic regime, the upwind configuration is not affected. Only a much
more detailed parametric study will be able to test these two possibilities.
  
\subsection{Physical properties of the critical solutions}

The solutions of this model, in particular Fig. 7, illustrate nicely the 
physical process of transferring electromagnetic Poynting energy flux and 
enthalpy to directed kinetic energy flux of the flow in order to accelerate 
a disk wind and then form a jet along the symmetry axis of the system. 
Thus, the analysis of the previous section is interesting in the sense that it 
reveals the driving mechanisms of the outflow. The poloidal kinetic energy 
is negligible at the disk level. It then increases rather sharply up to the 
region of the SMSS and Alfv\'en surfaces.  
This increase is at the expense of both, the enthalpy and the electromagnetic 
Poynting energy flux (see, Fig. 7). The poloidal velocity is directed  basically  
in the radial direction (Figs. 5 and 6), i.e., here part of the random thermal 
energy together with a part of the electromagnetic energy are mostly transformed  
to directed wind expansion.  
Downstream of the Alfv\'en surface it is mainly the Poynting energy flux that 
is effectively transformed into kinetic energy directed along the 
rotational axis, till the FMSS is encountered. After the FMSS, the 
flow has already reached the maximum speed available from the total energy E, 
which is also approximately equal to the initial electromagnetic Poynting energy 
flux. Then, the acceleration asymptotically stops.  
Despite the fact that most of the acceleration to high speeds is apparently of 
magnetic origin, the role of the 
polytropic index and thus of the initial thermal acceleration may not  
be negligible, in particular in the region before the SMSS.  
For example, in the case of Fig. 4 where the flow is exactly adiabatic and  
$\gamma=5/3$, the critical solution achieves only a very small axial component 
of the velocity which is twice the axial velocity on the equatorial plane.   
In the quasi-isothermal case of models I and II where $\gamma=1.05$, the 
maximum
velocity is 1000, higher than the equatorial one (Fig. 5). As a matter of fact, 
this last case is closer to the one analyzed in Li (1995) and Ferreira (1997)  
where the gas is isothermal up to the first critical surface and then it is 
taken 
to be cold afterwards, wherein the pressure has sufficiently dropped. 
However, another possibility is that the low terminal speed obtained in the 
adiabatic case of Fig. 4 could be due to the 
lower value of the rotation parameter $\lambda^2$ which is $\sim 2.8$ in the 
adiabatic case of Fig. 4, as opposed to values 
$\sim 137$ and 136 in models I and 
II and similarly for the case examined in Ferreira (1997).   

When the gas has reached a high speed along the $z$-axis, its inertia causes 
it to lag behind the rotation of the field line and the field is wound up, as 
shown in Fig. 6, resulting to a highly twisted magnetic field.  Consequently, 
the strong curvature force of this predominantly azimuthal magnetic 
field towards the $z$-axis, causes the poloidal field to collimate. Initially 
the field is flaring away from the rotation axis but the curvature force bends 
the poloidal field lines toward the rotation axis. The azimuthal velocity 
peaks around the Alfv\'en point which is at a height $z= 3.5$ and a 
cylindrical distance  $\varpi=5.8$ times the starting distance 
$\varpi_o$ in model I. Beyond the Alfv\'en point the rotation drops in 
accordance to angular momentum conservation and thus the centrifugal force 
becomes negligible. Then, the strong inwards curvature force of the twisted 
field, wins, over the weak outwards centrifugal force and gas pressure gradient 
with the result that the lines are bent and eventually collapse towards 
the rotation axis. 

It is interesting that this feature of the collapse of the outflow towards the 
rotation axis which appears in cold models (BP82) and models that do not cross 
the FMSS (Li 1995, Ferreira 1997), is also preserved in the present hot model 
where also all critical points are crossed. This result seems to indicate the 
rather dominant role of the magnetic hoop stress in radially self-similar 
models, contrary to what happens in meridionally self-similar models wherein   
the structure becomes asymptotically cylindrical (Trussoni et al. 1997, 
Sauty et al. 1999, Vlahakis \& Tsinganos 1999). 
 
It is worth to clarify for a moment the term  ``disk-wind'' that we used 
in this study. 
By that term we simply intend to indicate that we describe an outflow from a 
disk-like structure accreting onto a central gravitational object. 
Thus, the flow starts at some angle $\theta_o$ above or at the equatorial 
plane of the disk, as opposed to a ``stellar'' wind flow that starts radially 
above or at a spherical or quasi-spherical source. 
Needless to say that a consistent solution of the accreting part of the flow 
would be required for a consistent solution of the inflow-outflow structure 
in the case of a disk-wind. However, such a complete undertaking is beyond the 
scope of the present paper which only intends to emphasize the possibility to 
construct complete steady self-similar solutions for the wind crossing all 
critical points. 

To make such a connection between the disk and the outflow, in the spirit of 
BP82, Li (1995) and Ferreira (1997), the first step would be to see how our 
parameters may fall into the range of parameters considered by those models. 
For that purpose, in Eqs. (\ref{BPk}) - (\ref{BPxii}) we have made a 
correspondance between our parameters and those used by BP82.   
Thus, in the ``standard" solution of BP82 the parameters are:  
$\kappa_{\rm BP}= 0.03$, $\lambda_{\rm BP}
= 30$ and $\xi'_o = 1.58$ corresponding to a 
launching 
angle of the jet at the disk $\psi_o \approx 32^{\circ}< 60^{\circ}$.   
In our case, we find $\kappa_{\rm BP}\approx 0.13$, $\lambda_{\rm BP}\approx 
14.57$,  for both, model I and model II. We also 
have  
${\xi'_o}_{\rm BP}=\cot \psi_o=0.425$ ($\psi_o = 67^{\circ}$) for model I and   
${\xi'_o}_{\rm BP}=\cot \psi_o=0.675$ ($\psi_o = 56^{\circ}$) for model II, in 
the 
BP82 notation. We note that the values of $\kappa_{\rm BP}, \lambda_{\rm BP}$ 
are close in 
BP82 and the present model.
However, the 
value of the launching angle $\psi_o$  is $> 60^{\circ}$ in our model I because 
of the 
additional thermal 
driving of the outflow at the disk level, contrary to the cold model of BP82 
where $\psi_o \approx 32^{\circ} < 60^{\circ}$. 
In summary, our models I and II occupy in the space of $\kappa_{\rm BP}$ and
$\lambda_{\rm BP}$,  roughly the same domain as in BP82 
(cf. Fig. 2 in BP82). 
The only difference is in the value of the launching 
angle $\psi_o$ which can be 
larger in the present hot model, as expected.  
 These values are within the range of the allowed parameters in the 
($\kappa_{\rm BP}$, $\lambda_{\rm BP}$) space also in the analysis   
of Li (1995, cf. Fig. 3) provided that the magnetic diffusivity is of order 
one. 
 Note also that model II with $x=0.7525$  
 corresponds to an ejection index in the notation of Ferreira (1997) 
$\xi=2x-3/2=0.005$.

\subsection{Summary}

In this paper we have extended the classical work of Blandford and Payne 
(1982), mainly by showing via examples for the first time that a solution 
passing through all MHD critical points can indeed be constructed. 

As is well known, the  FMSS plays the role 
of the MHD signal horizon such that in an outflow crossing this MHD 
horizon all perturbations which the outflow may encounter are convected 
downstream by the superfast outflow and so the steady state solution is 
maintained. In other words, the outflow interior to the FMSS is causally 
disconnected and protected against any conditions it may encounter in the 
interstellar or intergalactic medium towards which the jet propagates 
after it is launched by magnetocentrifugal forces from the surface of an 
accretion disk.  

Unlike other analytical models which produce asymptotically 
cylindrically collimated outflows (Sauty \& Tsinganos 1994, Trussoni et al. 1997, 
VT98, Sauty et al. 1999, Vlahakis \& Tsinganos 1999), this class of 
radially 
self-similar models cannot continue to infinity but it has to be stopped 
downstream of the FMSS and matched via a MHD shock to a subfast outflow that 
mixes with the interstellar medium (Gomez de Castro \& Pudritz 1993).  
This shock can connect the present solutions 
to some breeze, subAlfv\'en or subslow magnetosonic branch perhaps also 
preserving the self-similarity. 
 
Thus, the main difference here with previous results presented in the 
literature is that the asymptotic part of the present solutions 
is causally disconnected from the source and hence any perturbation downstream 
of the superfast transition cannot affect the whole structure of the steady 
outflow.

This task of matching the present solutions with a downstream shock 
however remains a challenge for future studies,  
together with a (time-consuming) more extended parametric analysis and also 
a correct matching of the ideal MHD outflow solutions with an inflow in 
a non-ideal accretion disk (Ferreira 1997). 

\section*{Acknowledgments}

This research has been supported in part by a bilateral agreement between 
Greece and France (program Platon) and a NATO collaborative 
research grant between Greece, Italy and Russia. E.T. acknowledges the 
hospitality of the Observatoire de Paris and of the Department of Physics 
of the University of Crete. We wish to thank M. Micono for the 
information on the last data on protostellar jets and an anonymous referee 
for his comments which resulted in a better presentation of the paper.

\section*{References} 

   \refitem Bardeen J. M., Berger B. K., 1978, ApJ, 221, 105

   \refitem Blandford R. D., Payne D. G., 1982, MNRAS, 199, 883 (BP82)

   \refitem Bogovalov S. V., 1994, MNRAS, 270, 721
 
   \refitem Bogovalov S. V., 1996, MNRAS, 280, 39
   
   \refitem Bogovalov S. V., 1997, A\&A, 323, 634   

   \refitem Bogovalov S. V., Tsinganos K., 1999, A\&A, 305, 211 
 
   \refitem Burrows C. J., Stapelfeldt K. R., Watson A. M., et al., 1996, ApJ,   
      473, 451 

   \refitem Cabrit S., Edwards S., Strom S. E., Strom K. M., 1990, ApJ, 
      354, 687

   \refitem Cabrit S., Andre P., 1991, ApJ, 379, L25

   \refitem Contopoulos J., Lovelace R. V. E., 1994, ApJ, 429, 139 (CL94) 

   \refitem Contopoulos J., 1995, ApJ, 450, 616

    \refitem Ferreira J., 1997, A\&A, 319, 340

   \refitem Ferreira J., Pelletier G., 1995, A\&A, 295, 807

   \refitem Gomez de Castro A. I., Pudritz R. E., 1993, ApJ, 409, 748 

   \refitem  Habbal S. R., Tsinganos K., 1983, J. Geoph. Res., 88(A3), 1965
   
   \refitem Hartigan P., Edwards S., Ghandour L., 1995, ApJ, 452, 736

   \refitem K\"{o}nigl A., Pudritz R. E., 2000, in V. Manning, A. Boss, S.       
             Russel, eds, Protostars and Planets IV, University 
             of Arizona Press (astro-ph 9903168)

    \refitem Lery T., Henriksen R. N., Fiege J., 1999, A\&A,  350, 254

   \refitem Li Z-Y., 1995, ApJ, 444, 848

   \refitem Li Z-Y., 1996, ApJ, 465, 855
 
   \refitem Micono M., Massaglia S., Bodo G., Rossi P., Ferrari A.,
               1998, A\&A, 333, 1001
 
    \refitem Ostriker E., 1997, ApJ, 486, 291

    \refitem Ouyed R., Pudritz R. E., 1997, ApJ, 482, 712

   \refitem Padgett D., Brandner W., Stapelfeldt K., Strom S., 
     Tereby S., Koerner D., 1999, AJ, in press (astro-ph 9902101)

    \refitem Parker E. N., 1958, ApJ, 128, 664

    \refitem Ray T. P., 1996, in K. Tsinganos, ed, Solar and Astrophysical MHD  
      Flows,  Kluwer Academic Publishers, 539

     \refitem   Ray T. P., Muxlow T. W. B., Axon D. J., Brown A., Corcoran D.,
     Dyson J., Mundt R., 1997, Nature, 385, 415

      \refitem Ray T. P., 1998, in S. 
        Massaglia, G. Bodo, eds, Astrophysical jets: Open problems, 
        Gordon and Breach Science Publishers, 173

   \refitem Sauty C., Tsinganos K., 1994, A\&A, 287, 893  

   \refitem Sauty C., Tsinganos K., Trussoni E., 1999, A\&A, 348, 327 

   \refitem Spruit H. C., 1996, in R.A.M. Wijers et al., eds, Evolutionary        
             Processes in Binary Stars, Kluwer Academic Publishers, 249

   \refitem Trussoni E., Tsinganos K., Sauty C., 1997, A\&A, 325, 1099

   \refitem Tsinganos K., 1982, ApJ, 252, 775

   \refitem Tsinganos K., Sauty C., 1992, A\&A, 257, 790

   \refitem Tsinganos K., Sauty C., Surlantzis G., Trussoni E., 
       Contopoulos J., 1996, MNRAS, 283, 811

   \refitem Vlahakis N., Tsinganos K., 1997, MNRAS, 292, 591

   \refitem Vlahakis N., 1998,  Analytical Modeling of Cosmic
      Winds and Jets, PhD thesis, University of Crete, Heraklion

   \refitem Vlahakis N., Tsinganos K., 1998, MNRAS, 298, 777 (VT98)

   \refitem Vlahakis N., Tsinganos K., 1999, MNRAS, 307, 279
 
   \refitem Weber E.J., Davis L.J., 1967, ApJ, 148, 217
\newpage
\section*{Appendix}
\appendix
The two first order differential equations for  $G(\theta)$, $M(\theta)$ 
governing the present class of solutions are:
\begin{equation}\label{G'}
\frac{dG^2}{d\theta}=
\frac{2 G^2 \cos \psi}{\sin\theta \cos\left(\psi+\theta\right)}
\,,
\end{equation}
 
\begin{eqnarray}\label{M'}
\begin{array}{l}	
\displaystyle \frac{dM^2}{d\theta}=
-2\displaystyle \frac{\sin \left(\psi+\theta\right)}{\cos 
\left(\psi+\theta\right)} \left\{
-\displaystyle \frac{\kappa^2 \sin \theta}{ G} - \mu \left(x-2 \right) M^{4-2 
\gamma} +
\right. 
\\  \\
\left.
\displaystyle \frac{M^4}{G^4}\left(1-M^2\right) 
\displaystyle \frac{\cos\psi\sin\theta}{\sin \left(\psi+\theta\right)} -
\displaystyle \frac{M^4}{G^4}\left(x-2 \right) \displaystyle \frac{\sin ^2 
\theta}
{\cos^2\left(\psi+\theta\right)}-
\right.
\\ \\
\left.
\displaystyle \lambda^2 \displaystyle \frac{M^4}{G^2} \left(x-2 
\right) 
\left(\displaystyle \frac{1-G^2 }{ 1-M^2 } \right)^2 +
\displaystyle \lambda^2 \displaystyle \frac{M^2}{G^2} 
\frac{ G^4-M^2 }{ 1-M^2 }-
\right.
\\ \\
\left.
\displaystyle \lambda^2\displaystyle 
\frac{\cos\psi}{\sin\theta\sin 
\left(\psi+\theta\right)}
\displaystyle \frac{ \left(2M^2-1\right)G^4-M^4}{G^2\left(1-M^2\right)}
\right\} \times
\\ \\
\left\{
\gamma \mu\left(1-M^2\right)M^{-2 \gamma} 
-2\displaystyle \lambda^2 \displaystyle \frac{M^2}{G^2} 
\left(\displaystyle \frac{1-G^2 }{ 1-M^2 } \right)^2 +
\right.
\\ \\
\left.
2 \displaystyle \frac{M^4\sin^2 \theta}{G^4} 
\left(1-\displaystyle \frac{1}{M^2 \cos^2\left(\psi+\theta\right)}\right)
\right\}^{-1} \,.
\end{array}
\end{eqnarray}
\\
In the above two equations $\psi(\theta)$ is given by the Bernoulli integral:
\begin{eqnarray}\label{varphi}
\begin{array}{l}	
\psi=\pi-\theta \mp 
\arctan{}\left\{
\displaystyle \frac{G^4}{M^4 \sin ^2 \theta} \left[
2 \epsilon -\displaystyle \frac{\gamma \mu}{\left(\gamma-1\right)M^{2 
\left(\gamma-1\right)}}+
\right. \right.\\ \\
\left.  \left.
\displaystyle \frac{2 \kappa^2 \sin \theta}{ G} -\displaystyle \lambda^2
\left(
\displaystyle \frac{\left(G^2-M^2\right)^2}{G^2\left(1-M^2\right)^2}+2
\displaystyle \frac{ 1-G^2 }{ 1-M^2 } \right) \right] -1 \right\}  ^ {1/2} \,.
\end{array}
\end{eqnarray}
with the upper sign corresponding to the outflow case ($V_{r}>0$).\\

\noindent
On the Alfv\'en conical surface for $\theta \rightarrow \theta_\star$ we   
have
$$
\left(\displaystyle \frac{1-G^2}{1-M^2}\right)_{\star}=
\displaystyle \frac{2 \cos \psi_{\star} }{p_{\star} \sin \theta_{\star} 
\cos \left( \psi_{\star}+\theta_{\star}\right)} 
\,,
$$
where $p_{\star}$ is the slope of the square of the Alfv\'en number. Then from
Eq. (A2) we get the following third degree polynomial
for $p_{\star}$:
 
\begin{eqnarray}\label{alfven}
\begin{array}{l}
\left(x-2\right) \left(\displaystyle  4 \lambda^2+
p_{\star}^2 \sin^2 \theta_{\star} \right) \tan ^2
\left( \psi_{\star}+\theta_{\star}\right)+
\\ \\
\left(p_{\star}^3 \sin^2 \theta_{\star}+\displaystyle
4 \lambda^2 p_{\star}+  8 
\lambda^2\frac{\left(x-2\right)}
{\tan \theta_{\star}}\right) \tan \left( 
\psi_{\star}+\theta_{\star}\right)+
\\ \\
\left(x-2\right) \left(\mu p_{\star}^2+p_{\star}^2 \sin^2 \theta_{\star}
+\displaystyle 4\lambda^2 \frac{1}{\tan ^2\theta_{\star}}\right)
+\displaystyle \kappa^2 p_{\star}^2 \sin \theta_{\star}
\\ \\
-\displaystyle 
\lambda^2 p_{\star}\left(
p_{\star}-\frac{4}{\tan \theta_{\star}}\right)=0
\,.
\end{array}
\end{eqnarray}

\end{document}